\documentstyle[12pt,epsfig]{article}
\newcommand{\bpi}{\mbox{\boldmath $\pi$}}
\newcommand{\btau}{\mbox{\boldmath $\tau$}}
\def\haf{\textstyle{1\over2}}
\def\minus{\mbox{$-$}}
\newcommand{\vx}{\vec{x}}
\newcommand{\vsig}{\vec{\sigma}}
\newcommand{\vnabla}{\vec{\nabla}}
\newcommand{\fpi}{f_{\pi}}
\newcommand{\mpi}{m_{\pi}}

\newskip\humongous \humongous=0pt plus 1000pt minus 1000pt

\newif\ifdtup

\begin{document}
\vspace*{0.6in}
\hfill \fbox{\parbox[t]{1.12in}{LA-UR-04-4905}}
\vspace*{0.0in}

\begin{center}

{\Large {\bf Charge-Symmetry-Breaking Three-Nucleon Forces}}

\vspace*{0.4in}
{\bf J.\ L.\ Friar} \\
{\it Theoretical Division,
Los Alamos National Laboratory \\
Los Alamos, NM  87545} \\

\vspace*{0.10in}
\vspace*{0.10in}

{\bf G.\ L.\ Payne}\\
{\it Department of Physics and Astronomy, University of Iowa\\
Iowa City, IA 52242}\\

\vspace*{0.10in}
\vspace*{0.10in}

{\bf U.\ van Kolck}\\
{\it Department of Physics,
University of Arizona\\
Tucson, AZ 85721} \\
and \\
{\it RIKEN-BNL Research Center,
Brookhaven National Laboratory\\
Upton, NY 11973}\\

\end{center}

\begin{abstract}
Leading-order three-nucleon forces that violate isospin symmetry are calculated
in Chiral Perturbation Theory. The effect of the charge-symmetry-breaking
three-nucleon force is investigated in the trinucleon systems using Faddeev
calculations. We find that the contribution of this force to the $^3$He -- $^3$H
binding-energy difference is given by $\Delta E_{\rm 3NF}^{\rm CSB} \simeq 5$
keV.
\end{abstract}

\pagebreak

\section{Introduction}

Isospin violation \cite{iv1,iv2,iv3} has recently been investigated in the
context of Chiral Perturbation Theory ($\chi$PT). This powerful technique
\cite{weinberg,BvK} casts the symmetries of QCD into effective Lagrangians that
are expressed in terms of pions and nucleons, which are the effective degrees of
freedom of nuclear physics. These Lagrangian building blocks can then be
combined in a systematic way to develop isospin-violating forces. Although most
of the forces resulting from this procedure were anticipated and developed using
phenomenological methods, new forces have also been found. It is the purpose of
this work to complement the earlier work on isospin-violating two-nucleon
\cite{iv,1loop,pig,FvK,III,map,bochum1,bochum2,csb2} forces by calculating the
leading-order three-nucleon isospin-violating force, which breaks charge
symmetry. In addition, we estimate the contribution of this force to the $^3$He
-- $^3$H binding-energy difference. This is the first such calculation of
isospin-violating three-nucleon forces
\footnote{While this manuscript was being
written, we learned of a similar investigation ---using, however,
a different choice of fields--- by Epelbaum, Mei{\ss}ner and
Palomar \cite{emp}, where there is no attempt of a calculation of
binding energies.}
and completes the $\chi$PT calculations of isospin violation ---both
charge-independence breaking (CIB) and charge-symmetry breaking (CSB)--- through
the first three orders.

Chiral Perturbation Theory is organized around power counting (dimensional
analysis \cite{ndpc}), which allows estimates of the sizes of various mechanisms
to be made in terms of the parameters and scales intrinsic to QCD. These scales
(using Weinberg power counting \cite{weinberg,iv,primer}) include the pion decay
constant, $f_\pi \sim$ 93 MeV, which sets the scale for pion emission or
absorption, the pion mass, $m_\pi$, which sets the scale for chiral-symmetry
breaking, the typical nucleon momentum, $Q \sim m_\pi$ (which also determines
the inverse correlation length in nuclei), and the characteristic QCD
bound-state scale, $\Lambda \sim m_\rho$, which is appropriate for heavy mesons,
nucleon resonances, etc. The latter states are frozen out and do not explicitly
appear in $\chi$PT, although their effect is present in the counter terms of the
effective interactions. The resulting field theory is a power series in
$Q/\Lambda$, and the number of implicit powers of $1/\Lambda$ (e.g., $n$) can be
used to label individual terms in the Lagrangian (viz., ${\cal L}^{(n)}$).  In
this way higher powers denote smaller terms, and this is a critical part of the
organizing principle of $\chi$PT. We note that power-counting estimates of sizes
are typically within a factor of 2-3 of the actual sizes.

The nucleon-nucleon correlation-length scale is not relevant for one-body
operators, occurs once for two-body operators, twice for three-body operators,
etc. Thus it is important to incorporate this mechanism into the power counting
if we wish to compare mechanisms with differing numbers of interacting nucleons.
In addition to adding the indexes ``$n$'' for each of the individual Lagrangians
(see above) that are used in a given calculation, we must add 2 for each loop
and $N$-1 for an $N$-body Feynman diagram in order to determine the effective
order $\Delta$ \cite{primer}. Thus three-nucleon forces with an order determined
by an index $n-2$ should be comparable in size in nuclei to one-body operators
(such as the kinetic energy) corresponding to an index $n$, or a two-body force
corresponding to an index $n-1$. (Notice that this accounting of the relative
sizes of few-body forces is different from Weinberg's \cite{weinberg,iv} by one
order, which reflects the fact that we are counting the nucleon mass $M_N$ as
$\Lambda$, rather than   $\Lambda^2/Q$.) This is the underlying reason why
$N$-body forces in nuclei get systematically smaller as $N$ increases (and this
makes nuclear physics tractable).

Isospin violation in nuclei arises from three distinct mechanisms. The first is
the up-down quark-mass difference, which dominates and makes the neutron heavier
than the proton. The second mechanism is hard electromagnetic (EM) interactions
at the quark level, which tries to make the proton heavier than the neutron.
This is also the mechanism that produces most of the pion-mass difference. The
final mechanisms are the soft-photon interactions (such as the Coulomb
interaction between protons) that dominate isospin violation in nuclei.

Direct comparison \cite{III,map} of the sizes of the EM and quark-mass terms
demonstrates that the EM terms (which contain a factor of $\alpha$, the
fine-structure constant) are roughly the same size as quark-mass terms that are
formally three orders smaller in the power counting. We adjust our power
counting accordingly and adopt the convenient mnemonic of adding 3 to the order
of the EM-induced isospin-violating Lagrangian when comparing sizes with
quark-mass-induced mechanisms. Henceforth our power counting for any EM-induced
interactions will contain this additional factor of 3.

Our prior work on isospin-violating nucleon-nucleon forces (both CSB and CIB) in
the context of Chiral Perturbation Theory with the $\Delta$-isobar integrated
out \cite{iv,1loop,pig,FvK,III,map}  identified the leading mechanisms for
isospin violation in nuclei. In the next Section we discuss their impact on the
$^3$He -- $^3$H binding-energy difference. In the following Section we calculate
the leading isospin-violating three-nucleon force and evaluate its contribution
to $^3$He -- $^3$H binding-energy difference. We then conclude.

\section{Various CSB Mechanisms}

We have shown \cite{iv,1loop,III,map} that the following ten mechanisms are
expected to contribute dominantly to CSB in nuclei:

$\bullet$ The mass difference of the proton and neutron,
$\delta M_{\rm N} = m_p- m_n < 0$

$\bullet$ The CSB nuclear kinetic energy

$\bullet$ The Coulomb potential between protons

$\bullet$ The Breit-interaction ($(v/c)^2$) corrections to the Coulomb
potential

$\bullet$ The CSB one-pion-exchange potential (OPEP)

$\bullet$ The CSB short-range nuclear potential

$\bullet$ The CSB two-pion-exchange potential incorporating the nucleon-mass
difference

$\bullet$ The Class IV CSB interactions (anti-symmetric in isospin coordinates
and with a spin-orbit-type spin-space dependence)

$\bullet$ An OPEP with Class IV isospin structure that vanishes in the
two-nucleon center-of-mass, but not in a three-nucleon system

$\bullet$ A two-pion-exchange three-nucleon force proportional to the
quark-mass-difference contribution, $\delta M_{\rm N}^{\rm qm}$, to the
nucleon-mass difference


\noindent
We will briefly discuss each of them in turn in the context of the $^3$He --
$^3$H binding-energy difference.

The first four mechanisms are fairly well-known. They include the two largest
mechanisms, and we start our discussion with them.

The mass (rest-energy) difference of the nucleons, $\delta M_{\rm N}$,
contributes to the $\chi$PT Lagrangian at order $n=1$. From a nuclear-physics
perspective it makes an uninteresting contribution to the mass difference of
$^3$He and $^3$H and is conventionally removed, leaving only a binding-energy
difference. Although $^3$H is heavier than $^3$He, this removal leads to $^3$He
being less bound than $^3$H by 764 keV, which is the target for all CSB
calculations in the three-nucleon systems. The nucleon-mass difference
nevertheless plays a non-trivial role in intermediate states where two protons
are converted to two neutrons (or {\it vice versa}) by exchanging pions. That
effect was recently treated in a systematic fashion \cite{map} by removing the
$\delta M_{\rm N}$ mass-difference term from the $\chi$PT Lagrangian. This
removes $\delta M_{\rm N}$ from asymptotic states and nuclear energies, but its
effect in intermediate states is compensated by the addition of new terms in the
Lagrangian that must be incorporated in any calculations. The resulting scheme
is much simpler to use than older techniques, and we will use it below.

The kinetic-energy difference between two protons and two neutrons caused by
their different masses corresponds to $n=3$ in power counting. In the trinucleon
systems this mechanism leads to a robust 14~keV contribution
\cite{brandenburg,av18,sasakawa,argonne,nogga,fgp} to the binding-energy
difference of $^3$He and $^3$H.

The Coulomb potential between two protons is the dominant CSB interaction in
nuclei. According to the way we bookkeep EM interactions this is an effect one
order down compared to the leading, isospin-conserving nucleon-nucleon force, so
it is effectively an $n=1$, or $\Delta=2$, term. This contribution has a nominal
size in terms of scales given by $E_{\rm C} \sim \alpha Q \sim$ 1~MeV, where
$\alpha$ is the fine-structure constant. In the trinucleon systems it has been
well studied over several decades and leads to a robust and dominant 648~keV
contribution to the 764~keV trinucleon binding-energy difference \cite{coulomb}.

Small EM contributions of relativistic order contained in the Breit interaction
(viz., the interaction between nucleon magnetic moments and between the currents
associated with moving protons) plus a smaller vacuum-polarization force appear
two orders down ($n=3$ or $\Delta=4$ in our power counting). In terms of scales
the relativistic contributions behave like $E_{\rm B} \sim \alpha Q^3/\Lambda^2
\sim$ 25~keV. Indeed, they lead to a robust
\cite{brandenburg,av18,sasakawa,argonne,nogga} 28~keV.

The effect of CSB on two-nucleon potentials is subsumed by the next four
mechanisms on the list above, of which three are Class III and one is Class IV.
We will discuss them separately.

Charge-symmetry breaking in the pion-nucleon coupling constants can lead to a
CSB OPEP that has nominal size $n=2$, which corresponds to $\Delta = 3$. Only an
upper limit of size 50~keV (with unknown sign) constrains this mechanism
\cite{III}. A conventional short-range interaction of undetermined strength
corresponding to size $\Delta = 3$ (and a nominal size of roughly 50~keV) is
also present \cite{1loop}. The last of the three Class III mechanisms is the
recently calculated two-pion-exchange potential that incorporates various
aspects of the nucleon-mass difference. It is of nominal order $n=3$ or $\Delta
= 4$ \cite{III}. Each of these three mechanisms contributes to the difference
between the pp force (with the EM interaction removed) and the nn force. At
present the only experimental information on this difference is contained in the
scattering-length difference. The resulting $a_{nn}\minus a_{pp}$
scattering-length difference \cite{iv2,iv3,1loop} of $-$1.5(5) fm is then
attributed to CSB in the three forces discussed above, which cannot be further
disentangled at the present time. (Of course, in principle these mechanisms
could be separated thanks to their different ranges, provided that the
nucleon-nucleon data is accurate enough.) This difference then produces a
contribution to the $^3$He -- $^3$H binding-energy difference of approximately
65(22) keV, a number that also appears to be robust\cite{argonne,nogga,csmodel}.

The Class IV two-nucleon CSB OPEP \cite{iv2,iv3,map} has a nominal $n=2$ or
$\Delta = 3$ size, which is suppressed by nearly an order of magnitude by
nature's fine tuning of $\delta M_{\rm N}$ to its physical value \cite{map}.
This type of force is further suppressed in the trinucleon bound states because
S-wave components of the wave function do not contribute to a spin-orbit force.
Although this force and a short-range force of order $n=4$ are formally part of
the the CSB two-nucleon potential, they make a negligible contribution to the
trinucleon binding-energy difference.

The last two mechanisms are three-body effects.

The recent study \cite{map} of Class IV CSB forces found a peculiar two-body
force that vanishes in the center-of-mass of two nucleons, but does not vanish
in a system of more than two nucleons. Although nominally of order $n=2$ (or
$\Delta = 3$), this force should be much smaller than that for three reasons.
The first reason is that this two-body force is constrained by kinematics to
vanish in the center-of-mass of those two nucleons. In addition, this force is
proportional to $\delta M_{\rm N}$, which results from the cancellation of the
separate quark-mass and EM contributions and has been fine-tuned by nature to a
rather small value (a factor of 5 smaller than the nominal value of the
power-counting estimate for the quark-mass part $\delta M_{\rm N}^{\rm qm}$ of
that mass difference, viz., $-$7 MeV). The final suppression is caused by
approximate SU(4) symmetry in the few-nucleon systems. The spin-isospin
dependence of the force is antisymmetric under the interchange of those
coordinates for the two nucleons, caused by a $(\btau_i \times \btau_j)_z$ type
of isospin dependence. The dominant component of the trinucleon wave function
($\sim$ 90\%) is the S-state (an SU(4) classification), which is completely
antisymmetric under that interchange. These symmetry considerations cause the
diagonal S-state matrix element of the force to vanish. The net result of these
suppressions is that this force should be much smaller than its nominal order
indicates (i.e., $\Delta = 3$) and is therefore very unlikely to be significant.

The remaining force is a three-nucleon force of nominal order $n=1$ or $\Delta
=3$ that originates in the chiral-symmetry-breaking properties of the quark-mass
difference. It has never before been calculated, and we turn to it in the next
Section.

\section{CSB Three-Nucleon Forces}

In this section we examine isospin-violating three-nucleon forces within Chiral
Perturbation Theory. Isospin-conserving three-nucleon forces have been derived
within this approach in Refs. \cite{vK,cs3bf}. We follow the same method here.
In particular, we ignore terms that cancel against recoil in the iteration of
the two-nucleon potential.

The field redefinition that we employed in Ref.~\cite{map} eliminated the
nucleon-mass difference in the free Lagrangian at the cost of additional
effective interactions proportional to powers of that mass difference. Only
Lagrangian terms that had explicit time derivatives generated additional terms.
Incorporating the results of that field redefinition through orders $n=0$ and
$n=1$ in the Lagrangian (including short-range two- and three-body terms) plus
several other terms from Ref.~\cite{iv} leads to the following terms that
contribute to isospin-violating three-nucleon forces at orders $n=1$ (CSB) and
$n=2$ (CIB), plus omitted terms that would contribute only to higher orders:
\begin{eqnarray}
{\cal L}_{\rm iv} &=&
\frac{\delta M_{\rm N}^{\rm qm}} {4 f_{\pi}^2} N^{\dagger}
\left[\btau\cdot\bpi \pi_3 + ((\btau \times \bpi)\, \times \bpi)_3 \right] N
- \haf \left(\delta m_\pi^2 -\delta M_{\rm N}^2\right) \, (\bpi^2 -\pi_3^2)
\nonumber \\
&&+\frac{\tilde{c}_2 \delta M_{\rm N}^2 + \bar{\beta}_1/4}{f_{\pi}^2} \,
N^{\dagger} (\bpi^2 - \pi_3^2) N
+ \ldots  \; .
\label{eqno(1)}
\end{eqnarray}
The first of these terms breaks charge symmetry, while the remaining two break
charge independence. We will focus here on the first term, which is the largest
of all ($n=1$). We discuss corrections at the end of this Section.

\begin{figure}[tb]
\centerline{\epsfig{file=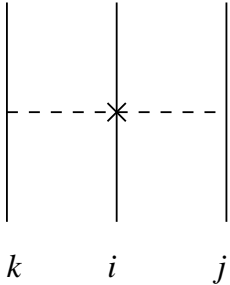,height=1.5in}}
\caption{Leading isospin-violating three-nucleon force $W_i$, which is
charge-symmetry breaking. A solid (dashed) line represents a nucleon (pion), and
the cross the interaction generated by the quark-mass difference component of
the nucleon-mass difference (first term in Eq. (\ref{eqno(1)})).}
\label{fig1}
\end{figure}

Using the lowest-order isospin-conserving Lagrangian
\begin{equation}
{\cal L}^{(0)}  = \frac{1}{2}[\dot{\bpi}^{2}-(\vnabla \bpi)^{2}
          -\mpi^{2}\bpi^{2}]
   + N^{\dagger}[i\partial_{0}-\frac{1}{4 \fpi^{2}} \btau \cdot
         (\bpi\times\dot{\bpi})]N +\frac{g_{A}}{2 \fpi}
 N^{\dagger}\vsig \cdot\vnabla(\btau \cdot \bpi)N +\ldots\, ,
\label{eqno(2)}
\end{equation}
a simple calculation along the lines of Ref. \cite{cs3bf} leads to the following
three-nucleon force corresponding to $\Delta = 3$. We define the total
three-nucleon force $W$ as
\begin{equation}
W = W_1 + W_2 + W_3 \, , \label{eqno(3)}
\end{equation}
where the subscript refers to the number of the nucleon that emits both pions
(the other two nucleons each absorb one of those pions), as shown in Fig.
\ref{fig1}. The expressions $W_i$ are symmetric under the interchange of
nucleons $j$ and $k$. We then find that
\begin{eqnarray}
W_1^{\rm CSB}
&=& -\frac{\delta M^{\rm qm}_{\rm N}\, g_A^2\, m^2_{\pi}}
{8\, f_{\pi}^4\, (4 \pi)^2} \left ( \;
\vsig_2 \cdot \hat{x}_{12}\; Y^{\prime} (|\vx_{12}|) \;
\vsig_3 \cdot \hat{x}_{13}\; Y^{\prime} (|\vx_{13}|) \; \right ) \nonumber \\
& &\times \left (\btau_1 \cdot \btau_2 \, \tau_3^{\rm z} +
          \btau_1 \cdot \btau_3 \, \tau_2^{\rm z} -
          \btau_2 \cdot \btau_3 \, \tau_1^{\rm z} \right ) \, , \label{eqno(4)}
\end{eqnarray}
where $\btau_i$ is the isospin operator for nucleon $i$, $\vsig_i$ is the spin
operator for nucleon $i$, $\vx_{ij}$ is the vector from nucleon $j$ to nucleon
$i$,
$$Y (x) = \exp{(-m_{\pi} x)}/(m_{\pi} x),$$
$g_A \cong 1.25$ is the axial-vector constant, and $\delta M_{\rm N}^{\rm qm}$
is the currently unknown quark-mass portion of the nucleon-mass difference.

The three-nucleon force in Eqs. (\ref{eqno(3)},\ref{eqno(4)}) is charge-symmetry
breaking. It is remarkable that this force appears formally at the same order as
the leading isospin-conserving  three-nucleon force \cite{vK,cs3bf} in the
theory without an explicit $\Delta$-isobar. That is, the factors of $Q$ and
$\Lambda$ are the same in both isospin-conserving and CSB three-nucleon forces
---although, of course, the CSB force is down by a factor of
$\varepsilon=(m_d-m_u)/(m_d+m_u) \sim 1/3$.

CSB is thus relatively large among three-nucleon effects, an unusual phenomenon.
It is agreed that three-nucleon effects (those not fixed by two-nucleon data)
provide about 1 MeV of the three-nucleon binding energies. The leading
three-nucleon forces contain relatively large sub-leading interactions (due to
effects of the $\Delta$-isobar), perhaps by a factor of 3 or so. Combining this
with a factor of $\varepsilon$, we could expect the CSB force to contribute as
much as 100 keV to the three-nucleon binding-energy difference. Indeed, using
the replacement $f_0^2 = (g_A \, m_\pi/2 \, f_{\pi})^2 / 4 \pi \cong 0.075$
(which is strictly valid only if the Goldberger-Treiman \cite{GT} relation is
exact, that is, to lowest order), Eq. (\ref{eqno(4)}) can be written as $-
2\delta M^{\rm qm}_{\rm N}\, f_0^4 / g_A^2$ times a dimensionless function of
coordinates (in units of $1/m_\pi$), spins, and isospins. Assuming that the
matrix elements of this function give numbers of order 1 and that $\delta M_{\rm
N}^{\rm qm}$ has its naive-dimensional-analysis value of $-$7 MeV, we arrive at
50 keV as an estimate for the size of the CSB three-nucleon force. This is
significant, but obviously it could differ from the actual value by a factor of
a few. The two sources of uncertainty in the size of this force are the values
of $\delta M_{\rm N}^{\rm qm}$ and of the dimensionless function above.

In order to better estimate the size of this CSB three-nucleon force, we have
implemented it in our Faddeev codes. The cutoff parameter in the TM$^{\prime}$
force \cite{cs3bf,TM} was adjusted slightly to produce the correct binding
energy for $^3$H when used in conjunction with the AV18 two-nucleon force
\cite{av18}. In any numerical calculation it is necessary to regulate the Yukawa
function, $Y (x)$ in Eq. (\ref{eqno(4)}), and this was done in a way that is
consistent with the Tucson-Melbourne force \cite{TM}. Perturbation theory was
then used to calculate the binding-energy difference of $^3$He and $^3$H. We
find
\begin{equation}
E_{\rm 3NF}^{\rm CSB} = 2 \times \left[-\frac{\delta M^{\rm qm}}{{\rm MeV}}
\right]\; {\rm keV} \, , \label{eqno(5)}
\end{equation}
which is about a factor 3 smaller than our estimate. Any other set of realistic
two- and three-nucleon forces should give similar results.

As mentioned above, the actual value of $\delta M_{\rm N}^{\rm qm}$ is
uncertain. It has been suggested \cite{npth,ddth} that it could be extracted
from pion-production experiments \cite{npexp,ddexp}, but it is unclear if this
can be achieved in the near future. It is likely to be smaller by a factor of a
few than the naive estimate, so $-$7 MeV is to be viewed as an overestimate.
Using $-$2.5 MeV for $\delta M_{\rm N}^{\rm qm}$, the contribution of our
three-nucleon force is listed in Table I together with all the other significant
contributions to the $^3$He -- $^3$H binding-energy difference that we have
discussed.

\begin{table}[b]
\centering
\caption{Contributions to the $^3$He -- $^3$H binding-energy difference in keV.
The Coulomb interaction and associated (relativistic) Breit-interaction
corrections dominate, while the CSB kinetic-energy difference (K.E.), the sum of
the short-range two-body CSB force mechanisms, and the three-nucleon CSB force
(calculated here for the first time) all make significant contributions. (In the
three-nucleon force, we used $\delta M_{\rm N}^{\rm qm}= - 2.5$ MeV for
illustration.) ``Theory'' labels the sum of these mechanisms.}

\hspace{0.25in}

\begin{tabular}{|ccccc|cc|}
\hline
Coulomb& Breit& K.E.& Two-Body  &Three-body& Theory  &Experiment\\ \hline
   648 &   28 &  14 &   65(22)  & 5        & 760(22) &764       \\ \hline
\end{tabular}
\end{table}

The three-nucleon results are in agreement with experiment when the error bar
associated with the strong-interaction CSB strength is taken into account. This
conclusion is also consistent with the CSB results extracted in Ref. \cite{CSB}
for $A=6 - 10$.

\begin{figure}[tb]
\centerline{\epsfig{file=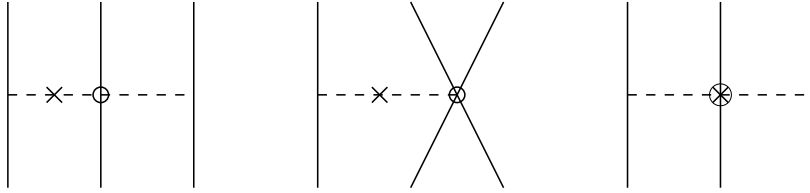,height=1.0in}}
\caption{Sub-leading isospin-violating three-nucleon forces from sub-leading
interactions. The cross represents the pion-mass difference (second term in Eq.
(\ref{eqno(1)})), while the circled cross stands for the sub-leading
isospin-violating seagull (third term in Eq. (\ref{eqno(1)})); a circle
represents an interaction from the sub-leading isospin-conserving Lagrangian.}
\label{fig2}
\end{figure}

There are, of course, other isospin-violating three-nucleon forces, but they are
higher order in our power counting and thus should be smaller. Some are
generated by sub-leading interactions, such as depicted in Fig. \ref{fig2}. The
second term in Eq. (\ref{eqno(1)}) reflects the additional amount that should be
added to the charged-pion mass (squared) in all pion propagators in
isospin-conserving three-nucleon forces,
such as the TM$^{\prime}$ force \cite{cs3bf}, which
comes from the sub-leading isospin-conserving Lagrangian ${\cal L}^{(1)}$.
The third term in Eq. (\ref{eqno(1)}) is an isospin-violating
contribution to the nucleon $\sigma$-term, often called $c_1$. It modifies the
three-nucleon force that is generated by charged-pion exchanges in that
interaction. Both of these modifications break, of course, charge independence,
but not charge symmetry.
These forces are transparently easy to implement, and we
refrain from writing explicit forms. They correspond to $n=2$ or $\Delta = 4$.

\begin{figure}[tb]
\centerline{\epsfig{file=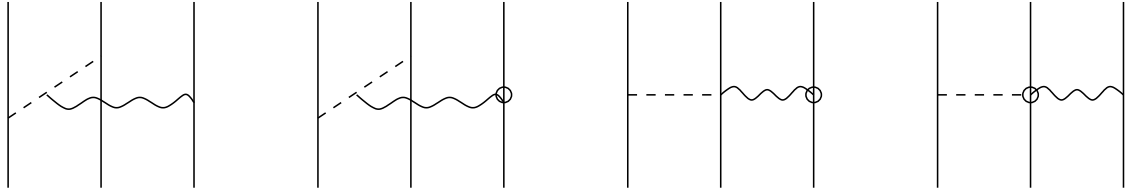,height=1.0in}}
\caption{Sub-leading isospin-violating three-nucleon forces from pion-photon
exchange. A wavy line represents a photon and a circle an interaction from the
gauging of the sub-leading isospin-conserving Lagrangian.}
\label{fig3}
\end{figure}

At the same sub-leading order there are also soft-EM forces, where a pion and a
photon are in the air at the same time, as in Fig. \ref{fig3}. While in flight
between two nucleons, a charged pion can exchange a photon with the third
nucleon. The photon couples either {\it i)} to the charge of the nucleon and the
energy of the pion, which in the nuclear environment is $Q^2/M_N$, or {\it ii)}
through the momentum of the pion and the magnetic moment of the nucleon, which
is a $1/M_N$ effect contained in the sub-leading Lagrangian ${\cal L}^{(1)}$. In
addition, there can be simultaneous emission of a photon and a charged pion by
one nucleon followed by their absorption on two other nucleons. This can happen
when the photon couples {\it i)} to the pion-nucleon vertex through the gauging
of the axial-vector coupling (third term in Eq. (\ref{eqno(2)})), and to the
nucleon magnetic moment, or {\it ii)} to the nucleon charge, and to the
pion-nucleon vertex through the gauging of the relativistic correction to the
pion-nucleon coupling contained in the sub-leading Lagrangian ${\cal L}^{(1)}$.
These mechanisms are formally $n=2$ in power counting (i.e., $-1 + 3$), and are
suppressed by one power of $Q/M_N$ compared to the CSB three-nucleon force we
calculated above. These three-nucleon forces therefore also correspond to
$\Delta = 4$. They break both charge independence and charge symmetry. Notice
that they are entirely determined by gauge and Galilean invariance in terms of
known parameters (the axial-vector coupling of the pion, the pion charge, the
nucleon charge and magnetic moment, and the pion and nucleon masses). Effects
from integrated-out resonances (most importantly the $\Delta$-isobar) only
appear one further order up. A subset of these EM effects has been calculated
before \cite{shuster,yang,yalcin}.

We expect that these uncalculated parts of the CSB force corresponding to higher
orders in the power counting contribute only a few keV or less, which is roughly
the level of uncertainty in the EM corrections discussed above.

\section{Conclusion}

After discussing various charge-symmetry breaking mechanisms in nuclei, we
derived the leading isospin-breaking three-nucleon force in Chiral Perturbation
Theory without an explicit $\Delta$-isobar, given by Eqs. (\ref{eqno(3)},
\ref{eqno(4)}). This force is charge-symmetry breaking and appears formally at
the same order as the leading isospin-conserving three-nucleon force. CSB could
thus be a relatively large three-nucleon effect. Its strength depends on $\delta
M_{\rm N}^{\rm qm}$, the contribution from the quark-mass difference to the
nucleon-mass difference. We therefore can directly tie QCD to a three-nucleon
effect. Unfortunately the actual value of $\delta M_{\rm N}^{\rm qm}$ has not
yet been determined in a model-independent way from low-energy data, nor from
lattice QCD.

We have also, for the first time, calculated the contribution of this force to
the $^3$He -- $^3$H binding-energy difference, given by Eq. (\ref{eqno(5)}).
Taking $\delta M_{\rm N}^{\rm qm}= - 2.5$ MeV for illustration, we find that 5
keV can be attributed to this force. This value is the same sign as the observed
difference, and is somewhat smaller in magnitude than expected from naive
dimensional analysis. As a consequence, it does not upset the agreement between
theory and experiment when the uncertainty in two-body effects is accounted for.

\section*{Acknowledgments}

UvK thanks RIKEN, Brookhaven National Laboratory and the U.S. Department of
Energy [DE-AC02-98CH10886] for providing the facilities essential for the
completion of this work. The work of JLF was performed under the auspices of the
U.S. Dept. of Energy. The work of GLP was supported in part by the DOE, while
that of UvK was supported in part by the DOE and the Alfred P. Sloan Foundation.

\end{document}